\documentclass[preprint,amsmath,amssymb,showpacs, aps]{revtex4}

\usepackage[dvips]{graphicx}

\begin{document}

\title{Fluctuations and universality in a catalysis model with long-range reactivity}

\author{C.~H. Chan}
\author{P.~A. Rikvold}
\email{prikvold@fsu.edu} 
\affiliation{Department of Physics, Florida State University, Tallahassee, Florida 32306-4350, USA}

\date{\today }

\begin{abstract}
The critical properties of the Ziff-Gulari-Barshad (ZGB) model
with the addition of long-range reactivity
[C. H. Chan and P. A. Rikvold, Phys. Rev. E \textbf{91}, 012103 (2015)] are further investigated.
The scaling behaviors of
the order parameter, susceptibility, and correlation length provide additional evidence that the universality class of the ZGB system
changes from the two-dimensional Ising class to the mean-field class with the
addition of even a weak long-range reactivity mechanism.\\
\end{abstract}


\maketitle

\section{Model and Motivation}
In 1986, Ziff, Gulari, and Barshad introduced a lattice-gas model (known as the ZGB model) to simulate the
nonequilibrium process of the oxidation of carbon monoxide
on a Pt catalyst \cite{Ziff1986}.
The catalyst surface is modeled as a $L\times L$ square lattice, and a random site is chosen $L^{2}$ times
in each time step (MCSS, Monte Carlo step per site).
Adsorption of the gas species can take place only if the chosen site is empty.
CO is adsorbed with probability $y$, which is proportional to its partial pressure in the supplied gas.
Adsorption of oxygen is attempted with probability $1-y$, but is successful only if a
randomly chosen nearest-neighbor site is also empty.
$\rm{O}_{2}$ then dissociates into two adsorbed O atoms.
After the adsorption, the adsorbed species checks its four nearest-neighbor sites.
If the opposite species is
found, the pair forms carbon dioxide ($\rm{CO}_{2}$) and leaves the surface immediately.
It has further been noticed in experiments
that adsorbed species can desorb from the surface without reacting, and that the desorption rate of carbon monoxide is much
higher than that of oxygen \cite{Ehsasi1989}.
Therefore, the desorption rate of CO ($k$) is often added to the model as a
second control parameter, in addition to the $\rm{CO}$ partial pressure ($y$) \cite{Brosilow1992,Tome1993}.
In this work, we choose desorption with probability $k$.
The desorption rate can be considered as a proxy for temperature as thermal fluctuations would make the
adsorbed particle break its bond with the catalyst surface more easily.

If the CO desorption rate ($k$) is not very high, simulations show that the catalyst surface can be in one of
three phases, depending on the value of the CO partial pressure ($y$): 1. CO poisoned phase, which has CO covering
most of the surface with some scattered empty sites. It is observed when $y$ is large.
2. O-poisoned phase, which has oxygen covering the whole surface when the CO
partial pressure ($y$) is sufficiently low (not observed in experiments).
3. mixed phase, in which the catalyst surface is covered by a mixture of O, CO, and empty sites. It is observed for
intermediate values of $y$.
The order parameter of the system is the CO coverage ($\theta_{\rm{CO}}$), which is defined as the proportion of the lattice
sites occupied by CO. A first-order phase
transition line separates the mixed phase and the CO poisoned phase. Similar to
an equilibrium lattice-gas system, moving along this first-order transition line in the phase
diagram eventually leads to a critical point. It has been found that this critical point belongs
to the two-dimensional equilibrium Ising universality class \cite{Tome1993}.

\begin{figure}[H]
\includegraphics[width=0.75\textwidth]{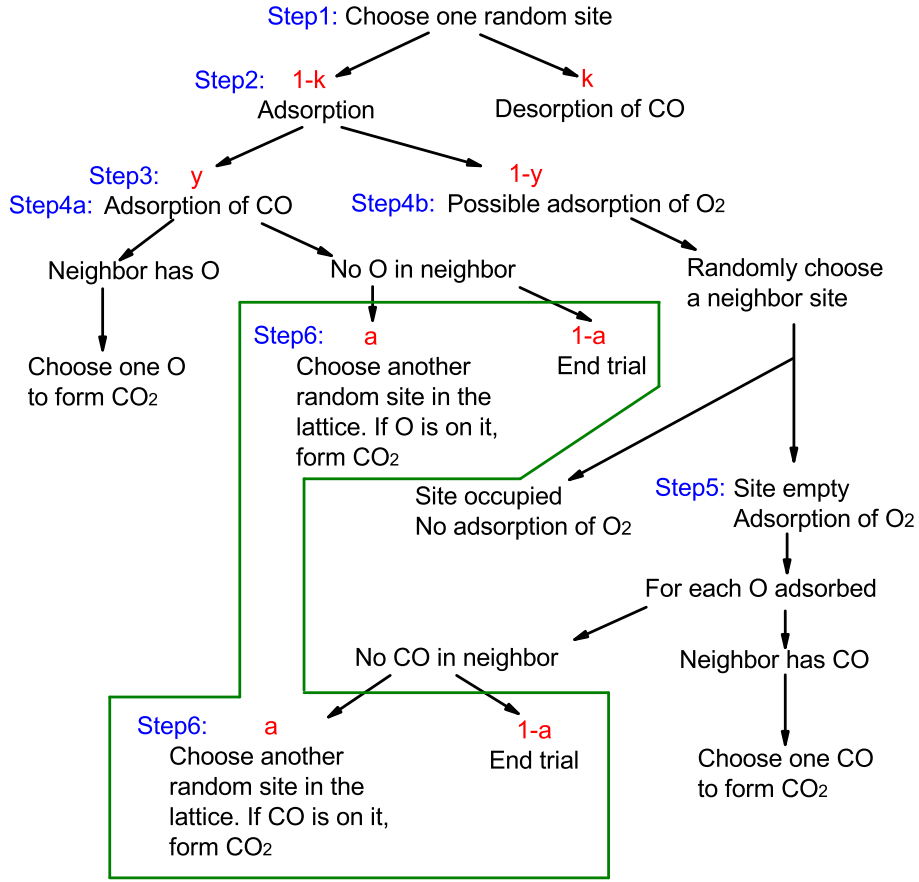}
\caption{Flow chart for the reaction process. The algorithm is based on that used in
\cite{Machado2005} for the ZGB model with CO desorption.
The framed region contains the added long-range reactivity of strength $a$. (From \cite{Chan2015}).
}
\label{fig_flow_chat_long_range}
\end{figure}

Recent studies of adding non-zero long-range interactions in an equilibrium Ising ferromagnet \cite{Nakada2011}
showed that the universality class of the critical point changed abruptly from Ising to mean-field.
Motivated by this result, we recently modified the ZGB
model by introducing an adjustable probability ($a$) that an O atom and a CO
molecule adsorbed far apart on the surface can react to form $\rm{CO_{2}}$ and desorb \cite{Chan2015}.
This move is attempted only
if the adsorbed species cannot find an opposite species within its four
nearest-neighbor sites. If the two sites
contain opposite species, $\rm{CO_{2}}$ will be formed and the two sites will become empty.
Details of the implementation are shown in Fig.~\ref{fig_flow_chat_long_range}.

While our recent work \cite{Chan2015} has shown that the universality class of
the critical point changes from the
Ising class to the mean-field class using fourth-order cumulants,
here we look at the fluctuations of the order
parameter (CO coverage) at the critical point through two quantities, the 'susceptibility' and the time average of
the absolute difference of the CO coverage from its mean value. The results further confirm the change of the universality
class of the critical point from Ising to mean-field upon introducing long-range reactivity.

\section{Results}

\begin{figure}[H]
\includegraphics[width=0.75\textwidth]{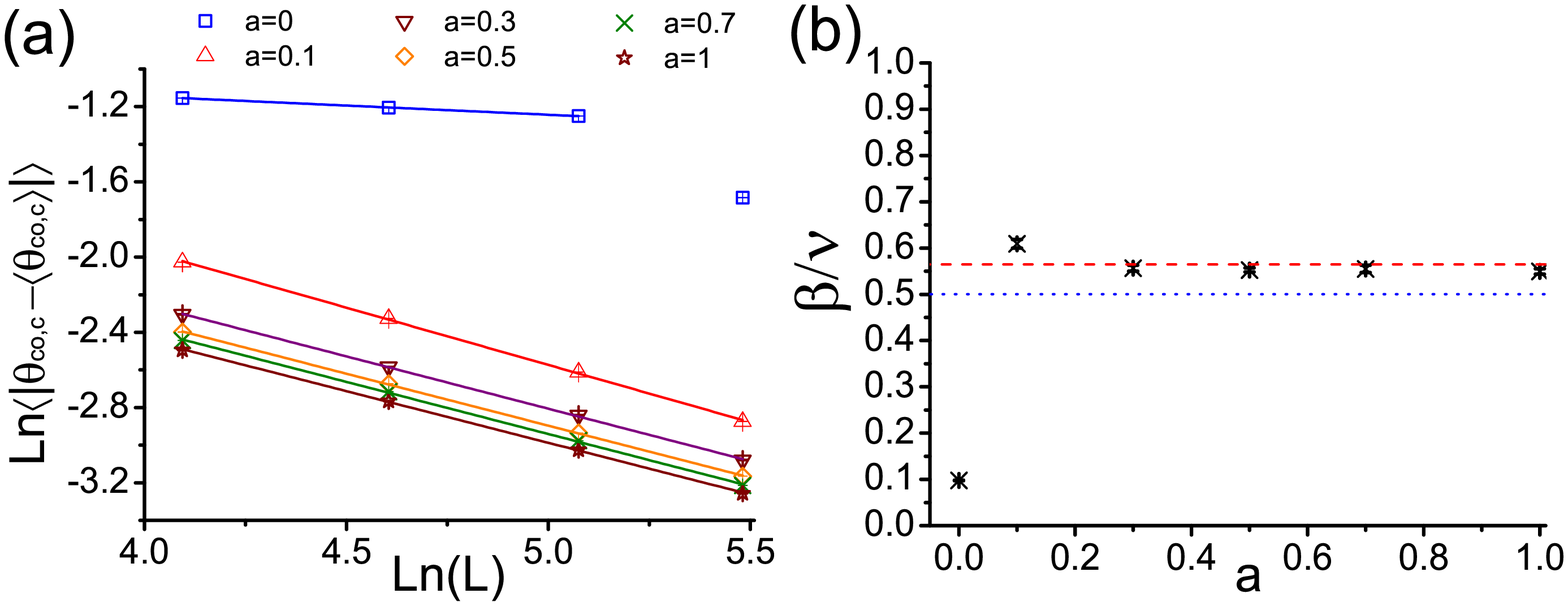}
\caption{
(a) Natural logarithm of the time average of the
difference of the CO coverage from its mean,
shown vs the natural logarithm of the size of the system
for $L=60$, $100$, $160$, and $240$. $ 10^{7}$ MCSS were used for
non-zero long-range reactivity ($a=0.1,0.3,0.5,0.7$ and $1$), and $ 10^{8}$ MCSS were used
without long-range reactivity ($a=0$), when the system was near the
corresponding critical point (e.g, the data point for $L=100$ was taken at the
$100/60$ cumulant crossing point). The negative of the slopes of these lines are the critical exponent ratio
$\beta/\nu$ and are plotted in (b). For $a=0$,
it gives $0.0977 \pm 0.0007$, which is close to the exact Ising value of $1/8=0.125$.
For $a>0$, the weighted average value is $0.564 \pm 0.008$ (dashed line),
which is close to the exact mean-field value of $1/2$ (dotted line).
}
\label{fig_ZGB_thesis_fakefluct_and_index}
\end{figure}

Figure~\ref{fig_ZGB_thesis_fakefluct_and_index}(a) shows a Ln-Ln scaling relation
 between the time average of the absolute difference of the CO
 coverage from its mean ($\langle|\theta_{\rm{CO,c}} - \langle\theta_{\rm{CO,c}}\rangle
|\rangle$) at the critical point, and the system size ($L$).
 The critical points were found in our recent work \cite{Chan2015},
 using cumulant crossing. The negatives of the slopes of these lines represent the
critical exponent ratio $\beta/\nu$ and are plotted in Fig.~\ref{fig_ZGB_thesis_fakefluct_and_index}(b).
Without long-range reactivity ($a=0$), we obtain $0.0977\pm 0.0007$, which is close to
the exact Ising value of $\beta=1/8$ and $\nu=1$. For the case
with long-range reactivity ($a>0$), the
weighted average is $0.564\pm 0.008$, consistent with the exact mean-field result of
$\beta=1/2$ and $\nu=1$.

\begin{figure}[H]
\includegraphics[width=0.75\textwidth]{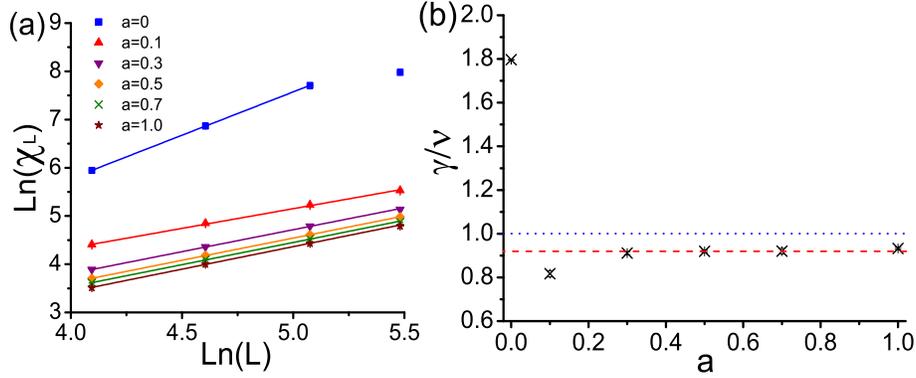}
\caption{
(a) Natural logarithm of the susceptibility
($\chi_{L}$), shown vs the natural logarithm of the system size ($L$)
for $L=60$, $100$, $160$, and $240$
near their corresponding critical points with $a=0$, $0.1$, $0.3$, $0.5$, $0.7$, and $1$.
$5\times10^{7}$ MCSS were used for $a>0$ and $10^{8}$ MCSS for
$a=0$. The slopes of the fitting lines for different long-range reactivities ($a$) are the critical
exponent ratio $\gamma/\nu$, plotted in (b). For the Ising case ($a=0$),
we obtain $\gamma/\nu = 1.7967 \pm 0.0001$, which is close to the exact Ising value of
$7/4$. For the mean-field case ($a>0$), we get a weighted average of
$\gamma/\nu = 0.92\pm 0.01$ (dashed line), which is close to the exact mean-field value of $1$ (dotted line).
}
\label{fig_ZGB_susceptibility_vs_L}
\end{figure}

We define a nonequilibrium analog of equilibrium magnetic susceptibility or fluid compressibility as
\begin{equation}\label{def_susceptibility}
\chi_{L}=L^{2}( \langle\theta^{2}_{{\rm{CO}},L}\rangle - \langle\theta_{{\rm{CO}},L}\rangle^{2}   )
\end{equation}
for a $L\times L$ system,
which for simplicity we call 'susceptibility.' It is well suited to measure the strong fluctuations of
the order parameter near the nonequilibrium critical point \cite{Machado2005}.
Figure.~\ref{fig_ZGB_susceptibility_vs_L}(a) shows
the natural logarithm of the susceptibility ($\chi_{L}$) vs the natural logarithm of the system size
($L$). The slopes of the fitting lines
represent the critical exponent ratio $\gamma/\nu$ \cite{Nakada2011} and are shown in
Fig.~\ref{fig_ZGB_susceptibility_vs_L}(b). For $a=0$, we obtain $\gamma/\nu = 1.7967 \pm 0.0001$, which
is close to the exact Ising value of $7/4$. For $a>0$, we get a weighted average of $\gamma/\nu = 0.92\pm
0.01$, which is close to the exact mean-field value of $1$.

\begin{figure}[H]
\includegraphics[width=0.75\textwidth]{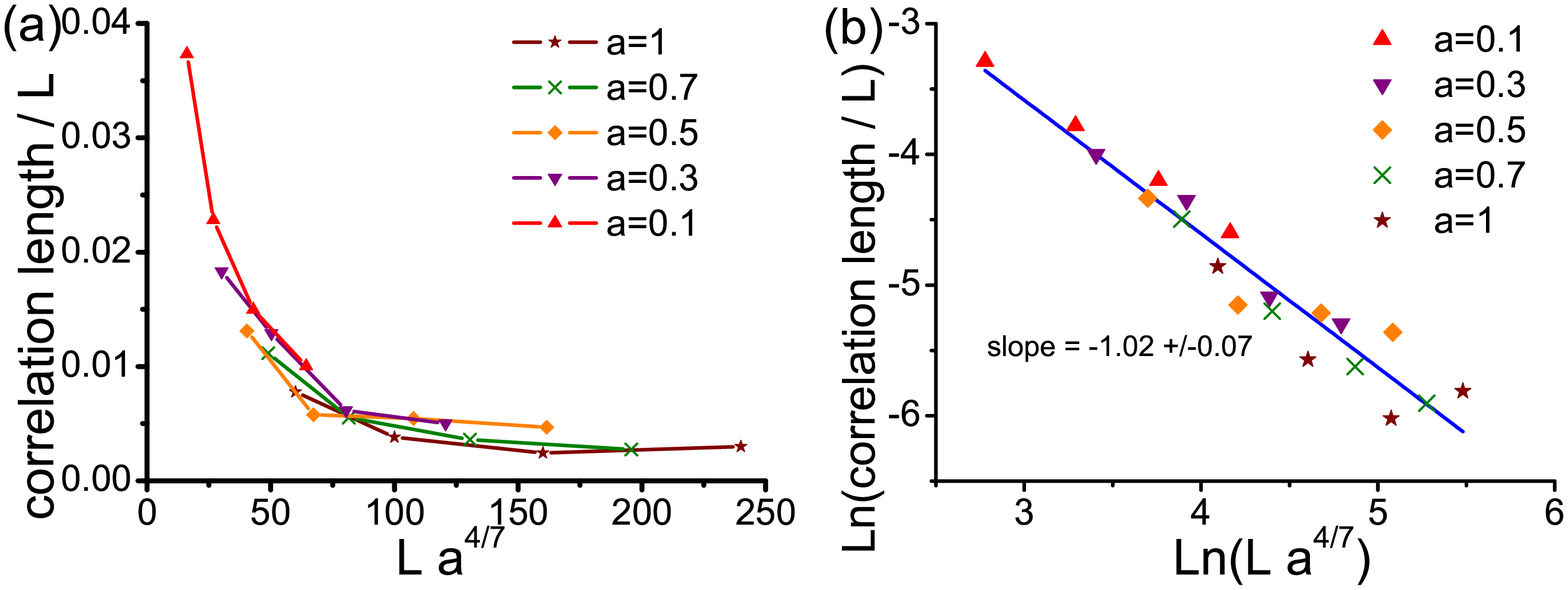}
\caption{
(a) Scaling plot of the correlation length at the
corresponding critical points with a scaling index of
$4/7$ for $a>0$, using $10^{7}$ MCSS. Scaling index values
of $0.435,0.448,0.499,0.45$ and $(4/7)/1.3$ were also tried but not shown,
as $4/7$ gives the best result, i.e., the data points fall most closely onto one curve.
A log-log plot of the results in (a)
are shown in (b). The data points lie along a straight
line of slope close to $-1$, which justifies the use of $4/7$ as
the scaling index.
}
\label{fig_correlation_scaling}
\end{figure}

We define the CO disconnected
correlation function as $c(r) = \langle \sigma_{i} \sigma_{j} \rangle$,
where $\sigma_{i}$ is $1$ if site $i$ is occupied by CO and is $0$ otherwise, $r$ is the distance
between site $i$ and site $j$, and the spatial
average is taken along the horizontal and vertical directions. The
critical correlation length was estimated by integration as
\begin{equation}\label{def_xi}
\xi(L) = \frac{\int^{L/2}_{0} [\langle c(r)\rangle - \langle c(L/2)\rangle] r dr}{\int^{L/2}_{0} [\langle c(r)\rangle - \langle c(L/2)\rangle] dr }
\end{equation}
as in \cite{Nakada2011} and \cite{Chan2015}.
%
In the study of an Ising model with long-range interactions of strength $\alpha$ \cite{Nakada2011},
Nakada \emph{et al.} suggested the scaling relation $\xi(L)=Lf(L\alpha^{\nu/\gamma})$.
They plotted $\xi(L)/L$ against $L\alpha^{\nu/\gamma}$
using the Ising value of $\nu/\gamma=4/7$ and verified that the data points all fell onto
one curve \cite{Nakada2011}.
When $\alpha$ was not too small, the log-log plot of the graph had
a slope of $-1$. Here we tried the same
procedure with our long-range reactivity $a$, also using $\nu/\gamma=4/7$, and
obtained Fig.~\ref{fig_correlation_scaling}(a).
Although the data points do not lie perfectly on one curve, in a Ln-Ln plot, we also
obtain a straight line with
slope close to $-1$, as shown in Fig.~\ref{fig_correlation_scaling}(b). Using $\nu/\gamma=1/1.7967$ as
obtained in Fig.~\ref{fig_ZGB_susceptibility_vs_L}, gives nearly exactly the same graphs, just
with the slope in (b) changed from $-1.02$ to $-1.03$.
 So it is justified to claim $\nu/\gamma=4/7$.

In our recent work \cite{Chan2015}, we also tried to obtain $\nu/\gamma$ by looking at the scaling relation between
the change in critical desorption rate with the change in
long-range reactivity, as well as
the scaling relation between change in critical
CO partial pressure with the change in long-range reactivity.
We obtained $k_{c,\infty}(a)-k_{c,\infty}(0) \propto  a^{0.448}$ and
$y_{c,\infty}(a)-y_{c,\infty}(0)  \propto  a^{0.499}$ and
expected that one of these values should be
 $\nu/\gamma$ of the $a=0$ case,
as we know that for the 2D equilibrium
Ising model with long-range interaction of strength $\alpha$, the change in critical temperature against the change in
long-range interaction strength $\alpha$ is known
to behave as $T_{c}(\alpha,\infty)-T_{c}(0,\infty) \propto \alpha^{\nu/\gamma}$ with
$\nu/\gamma$ obtained from the $a=0$ case, i.e. $4/7$ \cite{Nakada2011}.

The index values of $0.499$ and $0.448$, both deviate significantly from $4/7$.
We therefore suggested in \cite{Chan2015} that $a$ might not be linearly related to $\alpha$,
but rather as $a\propto \alpha^{1.3}$. However,
 Fig.~\ref{fig_correlation_scaling}(b) suggests that the relation for the equilibrium system, $\xi(L)=Lf(L\alpha^{\nu/\gamma})$,
 still holds for our model with $\alpha$ replaced by $a$, which means $a \propto \alpha$. Both
 Fig.~\ref{fig_ZGB_susceptibility_vs_L}(b)
 and Fig.~\ref{fig_correlation_scaling}(b) suggest that $\nu / \gamma  =4/7$ for $a=0$. Thus it seems that
 the indices obtained from
 $k_{c,\infty}(a)-k_{c,\infty}(0)$ vs $a$ and $y_{c,\infty}(a)-y_{c,\infty}(0)$ vs $a$ are not $\nu/\gamma$.
We currently have no satisfactory explanation for this result.

\section{Summary}
We present numerical evidence that the addition of non-zero long-range reactivity to the ZGB model
changes the universality class of the critical point from Ising to mean-field.
Estimates for the exponent ratios $\beta/\nu$ and $\gamma/\nu$
are obtained and compared to the value for $\gamma/\nu$ obtained in our previous published work \cite{Chan2015}.

This work was supported in part by NSF grant No. DMR-1104829.

\end{document}